\documentclass[12pt]{article}

\def\lsim{\mathrel{\rlap {\raise.5ex\hbox{$ < $}}
{\lower.5ex\hbox{$\sim$}}}}
\def\gsim{\mathrel{\rlap {\raise.5ex\hbox{$ > $}}
{\lower.5ex\hbox{$\sim$}}}}
\begin{document}

\begin{titlepage}
\begin{flushright}

hep-ph/xxxxx{\hskip.5cm}\\  
\end{flushright}
\begin{centering}
\vspace{.3in}
{\bf {Stabilized NMSSM without Domain Walls} }\\
\vspace{2 cm}
{C. PANAGIOTAKOPOULOS$^{1}$
 and K. TAMVAKIS$^{2}$}\\
\vskip 1cm
{$^1 $\it{Physics Division, School of Technology\\
Aristotle University of Thessaloniki, 54006 Thessaloniki, GREECE}\\
\vskip 1cm
{$^2$\it {Physics Department, University of Ioannina\\
45110 Ioannina, GREECE}}}\\

\vspace{1.5cm}
{\bf Abstract}\\
\end{centering}
\vspace{.1in}
We reconsider the Next to Minimal Supersymmetric Standard Model (NMSSM)
 as a natural solution to the $\mu$-problem and show
 that both the stability and the cosmological domain wall problems are 
 eliminated
  if we impose a ${\cal {Z}}_2$ $R$-symmetry 
on the non-renormalizable operators.

\vspace{1cm}
\begin{flushleft} 
September 1998
\end{flushleft}
\hrule width 6.7cm \vskip.1mm{\small \small}
 \end{titlepage}

The $N=1$ supersymmetric extension of the Standard Model provides a well
defined framework for the study of new physics beyond it \cite{NHK}. The low
energy data support the unification of gauge couplings in the supersymmetric
case in contrast to the standard case. The Minimal Supersymmetric extension
of the Standard Model (MSSM) is defined by promoting each standard field
into a superfield, doubling the higgs fields and imposing $R$-parity
conservation. The most viable scenario for the breaking of supersymmetry at
some low scale $m_{s}$, no larger than $\sim $ $1\ TeV,$ is the one based on
spontaneously broken supergravity. Although this scenario does not employ
purely gravitational forces but could require the appearance of gaugino
condensates in some hidden sector, it is usually referred to as
gravitationally induced supersymmetry breaking. The resulting broken theory,
independently of the details of the underlying high energy theory, contains
a number of \textit{soft} supersymmetry (susy) breaking terms proportional
to powers of the scale $m_{s}$. Probably the most attractive feature of the
MSSM is that it realizes a version of ``dimensional transmutation'' where
radiative corrections generate a new scale, namely the electroweak breaking
scale $M_{W}$. This is a highly desirable, but also non-trivial, property
that is equivalent to deriving $M_{W}$ from the supersymmetry breaking scale
as opposed to putting it by hand as an extra arbitrary parameter.
Unfortunately, a realistic utilization of radiative symmetry breaking \cite
{RSB} in MSSM requires the presence of the so called $\mu $-term coupling
directly the higgs fields $H_{1}$ and $H_{2}$, namely $\mu H_{1}H_{2}$, with
values of the theoretically arbitrary parameter $\mu $ close to $m_{s}$ or $%
M_{W}$. This nullifies all merits of radiative symmetry breaking since it
reintroduces an extra arbitrary scale from the back door. Of course, there
exist explanations for the values of the $\mu $-term, alas, all in extended
settings \cite{MU}.

At first glance, the most natural solution to the $\mu $-problem would be to
introduce a massless gauge singlet field $S$, coupled to the higgs fields as 
$\lambda SH_{1}H_{2}$, whose vacuum expectation value (vev) would turn out
to be of the order of the other scales floating around, namely $m_{s}$ and $%
M_{W}$. This leads to the simplest extension of the MSSM the so called
``Next to Minimal'' SSM or NMSSM \cite{NM} with a cubic (renormalizable)
superpotential 
\begin{equation}
{\mathcal{W}}_{ren}={\lambda }SH_{1}H_{2}+{\frac{\kappa }{3}}%
S^{3}+Y^{(u)}QU^{c}H_{1}+Y^{(d)}QD^{c}H_{2}+Y^{(e)}LE^{c}H_{2}.
\end{equation}

Unfortunately, the above scenario runs into difficulties. As can be readily
seen the NMSSM at the renormalizable level possesses a discrete
non-anomalous ${\mathcal{Z}}_{3}$ global symmetry under which all
superfields are multiplied by $e^{2\pi i/3}$. The discrete symmetry is
broken during the phase transition associated with the electroweak symmetry
breaking in the early universe and cosmologically dangerous domain walls are
produced. These walls would be harmless provided they disappear effectively
before nucleosynthesis which, roughly, requires the presence in the
effective potential of ${\mathcal{Z}}_{3}$-breaking terms of magnitude 
\[
\delta V\sim O(1\ MeV)^{4}\sim 10^{-12}\ GeV^{4}. 
\]
Such an estimate is not very different from the more elaborate one \cite{ASW}
\[
\delta V\sim 10^{-7}v^{3}M_{W}^{2}/M_{P}, 
\]
where $v$ is the scale of spontaneous breaking of the discrete symmetry and $%
M_{P}\simeq 1.2\times 10^{19}\ GeV\ $is the Planck mass. The above magnitude
of ${\mathcal{Z}}_{3}$-breaking seems to correspond to the presence in the
superpotential or in the K{\"{a}h}ler potential of ${\mathcal{Z}}_{3}$-
breaking operators suppressed by one inverse power of the Planck mass.
However, these ${\mathcal{Z}}_{3}$- breaking non-renormalizable terms
involving the singlet $S$ were shown \cite{ASW} to induce quadratically
divergent corrections\footnote{%
These non-renormalizable terms appear either as $D$-terms in the K{\"{a}}%
hler potential or as $F$-terms in the superpotential. The natural setting
for these interactions is $N=1$ Supergravity spontaneously broken by a set
of hidden sector fields.} which give rise to quadratically divergent
tadpoles for the singlet \cite{LAX}. Their generic form, cut-off at $M_{P}$,
is 
\begin{equation}
\xi m_{s}^{2}M_{P}(S+S^{*}),
\end{equation}
where $m_{s}$ is the scale of supersymmetry breaking in the visible sector.
The value of $\xi $ depends on the loop order of the associated graph (two
or three in this case) which, in turn, depends on the particular
non-renormalizable term that gives rise to the tadpole. Such terms lead to a
vev for the light singlet $S$ much larger than the electroweak scale. Thus,
it seems that the non-renormalizable terms that are able to make the walls
disappear before nucleosynthesis are the ones that destabilize the hierarchy.

The purpose of the present article is to address the two problems of domain
walls and destabilization that arise in the NMSSM and show that, despite the
impass that the previous arguments seem to indicate, there is a simple way
out rendering the model a viable solution to the $\mu $-problem. The crucial
observation is that due to the divergent tadpoles a ${\mathcal{Z}}_{3}$%
-breaking operator could have a much larger effect on the vacuum than its
dimension naively indicates. Thus, it is conceivable that non-renormalizable
terms suppressed by more than one inverse powers of $M_{P}$ are able to
generate linear terms in the effective potential which are strong enough to
eliminate the domain wall problem although, at the same time, they are too
weak to upset the gauge hierarchy. Clearly, it would be very helpful to
obtain a better understanding of both the symmetries that could be imposed
on the model and the magnitude of destabilization that the various
non-renormalizable operators generate.

The renormalizable part of the NMSSM superpotential (1) possesses the
following global symmetries: 
\[
U(1)_{B}:Q(\frac{1}{3}),\,U^{c}(-\frac{1}{3}),\,D^{c}(-\frac{1}{3}%
),\,L(0),\,E^{c}(0),\,H_{1}(0),\,H_{2}(0),\,S(0) 
\]
\[
U(1)_{L}:\,Q(0),\,U^{c}(0),\,D^{c}(0),\,L(1),\,E^{c}(-1),\,H_{1}(0),%
\,H_{2}(0),\,S(0) 
\]
\[
U(1)_{R}:\,Q(1),\,U^{c}(1),\,D^{c}(1),\,L(1),\,E^{c}(1),\,H_{1}(1),%
\,H_{2}(1),\,S(1) 
\]
(where in parenthesis is given the charge of the superfield under the
corresponding symmetry). The last $U(1)$ is an anomalous $R$-symmetry under
which the renormalizable superpotential ${\mathcal{W}}_{ren}$ has charge 3.
The soft trilinear susy-breaking terms break the continuous $R$-symmetry $%
U(1)_{R}$ down to its ${\mathcal{Z}}_{3}$ subgroup that we mentioned earlier
which, however, is not an $R$-symmetry. We see that the renormalizable part
of the model possesses a genuinely discrete symmetry whose spontaneous
breakdown produces domain walls.

Of cource, one does not have to impose all the above continuous symmetries
in order to obtain the renormalizable superpotential ${\mathcal{W}}_{ren}$
of the NMSSM. The same ${\mathcal{W}}_{ren}$ can be obtained if we impose a
discrete symmetry. There are various choices among which it is useful to
consider two interesting possibilities:

\textbf{{a) ${\mathcal{Z}}_{2}^{MP}\times {\mathcal{Z}}_{3}.$}} The \textit{%
matter parity} ${\mathcal{Z}}_{2}^{MP}$ is generated by 
\[
{\mathcal{Z}}_{2}^{MP}\,:\,(Q,U^{c},D^{c},L,E^{c})\rightarrow
-(Q,U^{c},D^{c},L,E^{c}),\ \ \,(H_{1},H_{2},S)\rightarrow (H_{1},H_{2},S) 
\]
and the ${\mathcal{Z}}_{3}$ symmetry by 
\[
{\mathcal{Z}}_{3}\,:\,(Q,U^{c},D^{c},L,E^{c},H_{1},H_{2},S)\rightarrow
e^{2\pi i/3}(Q,U^{c},D^{c},L,E^{c},H_{1},H_{2},S). 
\]
Note that ${\mathcal{Z}}_{3}\subset U(1)_{R},$ as already mentioned. Both ${%
\mathcal{Z}}_{2}^{MP}$ and ${\mathcal{Z}}_{3}$ are not $R$-symmetries (${%
\mathcal{W}}\rightarrow {\mathcal{W}}$).

\textbf{{b) ${\mathcal{Z}}_{2}^{MP}\times {\mathcal{Z}}_{4}^{(R)}.$}} The
matter parity ${\mathcal{Z}}_{2}^{MP}$ generator is defined as in the
previous case. The ${\mathcal{Z}}_{4}$ $R$-symmetry ${\mathcal{Z}}%
_{4}^{(R)}\subset U(1)_{R}$ generator is defined by

\[
{\mathcal{Z}}_{4}^{(R)}\,:\,(Q,U^{c},D^{c},L,E^{c},H_{1},H_{2},S)\rightarrow
i(Q,U^{c},D^{c},L,E^{c},H_{1},H_{2},S),\ \ {\mathcal{W}}\rightarrow -i{%
\mathcal{W}}. 
\]

Although it makes no difference which of the above symmetries are imposed on
the renormalizable superpotential, we should make sure that the ${\mathcal{Z}%
}_{3}$ symmetry, or any other symmetry containing it, is not a symmetry of
the non-renormalizable operators. If ${\mathcal{Z}}_{3}$ invariance is
imposed on the complete theory the domain walls will not disappear. In
contrast, the ${\mathcal{Z}}_{4}^{(R)}$ symmetry can be imposed on the
non-renormalizable operators and no domain walls associated with its
breaking will form because the soft susy-breaking terms break ${\mathcal{Z}}%
_{4}^{(R)}$ completely.

Let us now move to the other important issue that has to be addressed in the
presence of the gauge singlet superfield $S$, namely the destabilization of
the electroweak scale due to quadratically divergent tadpole diagrams
involving non-renormalizable operators which generate in the effective
action linear terms of the type (2). As mentioned, such terms lead to a vev
for the light singlet which, in general, is much larger than the electroweak
scale. Abel \cite{ABEL} has shown that \textit{the potentially harmful
non-renormalizable terms are either even superpotential terms or odd K{\"{a}}%
hler potential ones}. Such terms are easily avoided if we impose on the
non-renormalizable operators a ${\mathcal{Z}}_{2}$ $R$-symmetry ${\mathcal{Z}%
}_{2}^{(R)}$ under which the superpotential as well as all superfields flip
sign. This symmetry is a subgroup of both $U(1)_{R}$ and ${\mathcal{Z}}%
_{4}^{(R)}$. Therefore, one has the option of imposing on all operators a
symmetry ${\mathcal{Z}}_{2}^{MP}\times {\mathcal{Z}}_{4}^{(R)}$ or ${%
\mathcal{Z}}_{2}^{MP}\times {\mathcal{Z}}_{2}^{(R)}$ or just ${\mathcal{Z}}%
_{2}^{(R)}$ assuming in the last two cases that the renormalizable
superpotential has \textit{accidentally} a larger symmetry.

Notice that the non-renormalizable terms allowed by ${\mathcal{Z}}_{2}^{(R)}$
or ${\mathcal{Z}}_{4}^{(R)}$, although not harmful to the gauge hierarchy,
are still able to solve the ${\mathcal{Z}}_{3}$-domain wall problem since
they generate in the effective action through $n$-loop tadpole diagrams
linear terms of the form 
\[
\delta V\sim (16\pi ^{2})^{-n}m_{s}^{3}(S+S^{*}). 
\]
These terms are small to upset the gauge hierarchy but large enough to break
the ${\mathcal{Z}}_{3}$ symmetry and eliminate the domain wall problem. For
example, the presence of the term $S^{7}/M_{P}^{4}$ in the superpotential,
allowed by both symmetries ${\mathcal{Z}}_{2}^{(R)}$ and ${\mathcal{Z}}%
_{4}^{(R)}$, is able to generate at four loops such a harmless linear term,
as shown by Abel \cite{ABEL}.

Combining all the above we see that by adopting the renormalizable
superpotential (1) of the NMSSM and imposing on the non-renormalizable
operators just a ${\mathcal{Z}}_{2}$ $R$-symmetry ${\mathcal{Z}}_{2}^{(R)}$
we are able to solve both the cosmological and the stability problems of the
model\footnote{%
Incidentally notice that ${\mathcal{Z}}_2^{(R)}$ (or ${\mathcal{Z}}_4^{(R)}$)
eliminates all dimension five operators leading to fast proton decay.}. Thus,
NMSSM can be finally regarded as a solution to the $\mu $-problem of the
MSSM without invoking non-minimal K{\"{a}}hler potentials coupling directly
visible and hidden sector fields.

{\Large \textbf{Aknowledgements}}

C. P. was supported in part by CERN as a Corresponding Associate and by the
TMR network ``Beyond the Standard Model''. C. P. wishes to thank S. Abel for
a useful discussion. K. T. wishes to thank the TMR network ``Beyond the
Standard Model'' and the Greek Ministry of Science and Technology, through
its $\Pi $ENE$\Delta $ and research collaboration with CERN programs, for
travelling support. He also wishes to thank the CERN Theory Division for its
hospitality during the summer of 1998.

\end{document}